Experimental methods to measure thermal hysteresis of martensite-austenite transition in bulk β-Cu-Zn-Al composition gradient single crystals

O.Goisot[1,2], L.Porcar[2*], R.Haettel[2], H.vanLandeghem[1], F.Robaut[1], O.Robach[3], M.Verdier[1]
(1) Univ. Grenoble Alpes, CNRS, Grenoble INP, Institute of Engineering Univ. Grenoble Alpes, SIMaP UMR 5266, F-38000 Grenoble, France
(2) Univ. Grenoble Alpes, CNRS, Grenoble INP, Institute of Engineering Univ. Grenoble Alpes, Institut Néel UPR 2940 , F-38000 Grenoble, France
(3) Univ. Grenoble Alpes, CNRS, CEA, IRIG, MEM-NRSUMR 9001, Grenoble 38000, France

(*) corresponding author: laureline.porcar@neel.cnrs.fr

**Abstract** : Improvement of functional and structural fatigue endurance for applications of ferroelastic materials requires an optimization of their composition. We propose an experimental methodology to explore the model β-Cu-Zn-Al system. It is based on an original route to process bulk gradient composition single crystals to investigate fine variation of composition range coupled with local measurements of the austenite-martensite microstructure by light microscopy during the transformation. The latter method is compared with differential scanning calorimetry measurements. The methodology is applied to different narrow ranges (up to 4 at%) of composition of standard CuZnAl shape memory alloys where local fluctuations of hysteresis of 3 K are observed.

## Introduction

Martensitic phase transformations, which are first-order and displacive, can be used in various fields of application ranging from shape memory alloys actuators, energy, biomedical to structural damping applications [1]. These applications require forward and reverse martensitic transformation cycling. Extending materials fatigue life remains, in many cases, a major scientific and technological challenge. For example, in the case of realistic heat pump based on elasto-caloric materials, cycling must be above ten million cycles to guarantee one year of operation at 1 Hz. The problem lies in finding a material capable of withstanding such a high number of transformation cycles without losing its thermodynamic properties ($\Delta S$, $\Delta T_{adiabatic}$), known as functional fatigue, and without fracturing, known as structural fatigue. Both fatigue limits come from irreversible mechanism that entails reversibility and produce structural defects (dislocations) [2, 3]. Three main alloy families of shape memory alloys are usually considered for general applications: NiTi alloys, their derived multiferroic NiMnX compositions and the Cu-based alloys[4]. Cu-based SMAs have good mechanical properties and highly reversible thermoelastic transformation. They are good candidates for extending fatigue life [5].

The fatigue behavior and reversibility are dependent on the processing conditions and microstructure and we review here trends influencing the reversibility of the transformation in shape memory alloys.

- A slight change in composition (less than 1 at%) can shift transformation temperatures dramatically. Moreover, in ferroelastic metallic alloys, a significant effect of composition on thermal hysteresis width (from 20 K to a few K) has been measured in CuZnAl [6] and CuZnAu [7]. It has been demonstrated that materials exhibiting good crystallographic compatibility during the martensitic transition have high reversibility, which results in low thermal hysteresis and minimal broadening or shift of the hysteresis curves [8]. Thus, in ternary $Cu_{55-x}Au_xZn_{45}$ alloys [7], the variation in hysteresis width can reach 16 K for variations of 1 at% in x. The lowest hysteresis is less than 2 K, which naturally made this ternary alloy the material of choice for demonstrating super-compatibility based on the nonlinear crystallographic theory of martensite [9] and it has been successfully applied to specific NiTi alloys [10]. Certain compositions allow for a 'tolerant compatibility', where compatible stresses are sufficiently low to preserve reversibility as in CuZnAl [11, 12]. Optimizing alloys such as CuAuZn or NiTiCu to achieve exceptional fatigue resistance required investigations with very fine variations in composition (less than 1%) to find wells of minimum hysteresis [13], either by processing a very large number of samples or by combinatorial approach such as thin film composition gradient processing [13]. However, the development of thin films leads to problems related to size, defects (grain size) and intrinsic residual stress resulting from the deposition process, unlike the bulk combinatorial approach that we propose in this work.
- The austenitic phase is a stable β-phase at high temperatures and therefore metastable at room temperature. The ordering of austenite is a crucial parameter in the martensitic transition and the transformation temperatures are directly influenced by it [14]. Depending on the heat treatments applied (quenching speed, up-quenching, annealing), the crystallographic ordering of the cubic austenite phase may differ, especially on off-stoechiometric compositions from the ideal $L2_1$ ($Cu_2ZnAl$) composition [15].
- Grain size in poly-crystalline bulk sample is known to dramatically impact hysteresis and fatigue, being sites of strain incompatibilities, especially for the highly anisotropic elasticity of β-brass. Thermal hysteresis $\Delta T_{hyst}$ is inversely proportional to the grain [16].
- For sample size processed with one dimension and large free surface to volume ratio, such as wire or ribbons with oligocrystalline microstructure, fatigue lifetime is improved two orders of magnitude over poly-crystalline Cu-based alloys [16–18].

- Martensitic transformation is driven by a shear mechanism:, consequently, residual stresses and defects significantly influence the characteristic transformation temperatures. Moreover, obtaining a defect free surface is crucial, as surface imperfections can retain martensite a broader the transformation, as well as provide dislocations that can propagate into the bulk with cycling, thereby reducing the fatigue strength of the materials [19, 20].

In summary, numerous studies report the influence of processing conditions, structural defect density, ordering of the β-phase, microstructure on thermodynamic and reversibility of the ferroelastic transition in CuZnAl. However, the literature draws on data collected under a wide variety of processing and measurements conditions for CuZnAl alloys. For that purpose we propose an original route to produce controlled composition gradient in high-quality single- /oligo- crystals.

In this paper, we propose a methodology (sample processing and measurement method) for systematic measurements of possible variations in transformation temperatures and hysteresis. This is achieved in bulk single-crystal samples with controlled range of fine composition variation that have undergone identical processing and preparation conditions.

Since the samples exhibit concentration gradients, transformation temperatures can only be measured locally using spatially resolved methods. We propose here to measure the austenite / martensite transition during thermal cycling based on direct surface observation of the microstructures. We use polarized light microscopy (PLM). Indeed, PLM was used by [19], in conjunction with an unconventional calorimetric device (called thermomicroscope) to establish a quantitative link between various aspects of the transition and the evolution of the mesoscopic microstructures (~100µm of FOV).

In the first part, we detail our processing route for obtaining a controlled β-CuZnAl solid solution single crystal with defined composition gradients. In the second part, we describe our methodology for measuring characteristic temperatures and compare it to standard differential scanning calorimetry (DSC). The procedure is applied to compositions around the standard Cu68Zn16Al16 SMA [15, 21] where the influence of composition is presented and discussed.

## Processing of bulk single crystal with concentration gradients

In all samples, the chemical composition were measured by electron probe micro Analysis (EPMA Cameca SX-100) at 24 kV for optimal measurements of the K level of Zn with a regulated beam current of 65 nA. Whole surface sample were measured with grids of 100 µm steps along gradients, and 400 µm perpendicular to it.

Relocation of specific areas using the different characterization techniques (EPMA, PLM, EBSD) where achieved using nanoindent markers. Samples are cut with a precision micro cutting machine and polished down to 1 µm with diamond suspension. The finishing steps consist of removing the damaged layer by vibratory polishing with a colloidal suspension of 40nm silica particles for 16h.

Single crystals, both homogenous and with compositions gradients are processed:

- Homogeneous CuZnAl alloys were prepared by melting Cu, Zn, and Al elements with a purity of 4.5 N encapsulated in a quartz-glass capsule at low pressure (700 mbar of Ar) – around 10g load. The fusion was carried out at 1373 K in a muffle furnace, with cooling to 1173 K over 3 hours, followed by a holding time of 24 hours, then cooling in air to room temperature (approximately 50 K/min). This results in mm size large grains as observed in the literature [16]. Out of the 10 mm diameter ingot, 3mm single crystalline disks sample where cut by wire spark cutting with a chemical variation of +/- 0.15 at.% was measured.

For generating bulk concentration gradients two methods were used:

- A conventional method is based on diffusion couple bonding where two samples with different compositions are joined by applying a constant pressure in an inert atmosphere at high temperature [22]. We applied this method using 4.5 N purity elements and obtained a gradient sample by applying during 1 hour a stress of 1 MPa at 1123K in 1 bar (Ar+2% $H_2$) [23]. It enabled us to measure the apparent diffusion coefficients of Zn and Al as a function of diffusion annealing time and temperatures. The diffusion coefficient of zinc was measured for 1173 K. It was found to be $D_{Zn} = 7.10^{-7}$ cm²/s, which is consistent with the data reported for β-CuZn [24]. Obtaining a low composition gradient (1 at% over 10 mm) can be cumbersome with this method since it implies the processing and homogenization of the two end crystals.

- An original method has been developed, which principle is given in Figure 1 for sample Grad.1 and Grad.2. A 1.5 mm high Zn cylinder(A) is placed on a 10 mm high Cu79Al21 cylinder (A') so that the whole corresponds to a target composition (C) Cu69Zn14Al17. The cylinders were machined and then placed in a sealed quartz cylinder which was vacuum pumped to $10^{-5}$ mbars and filled with 700 mbar of Ar and heated to $T_{annealed}$ in a muffle furnace. For this composition, $T_{annealed}$ =1173 K is chosen well below the liquidus of the target composition(C), 1273 K, Figure 1b.
Initially, pure Zn melts first and the liquid phase dissolves the solid $Cu_{79}Al_{21}$. The liquid front propagates downwards until it reaches saturation at the equilibrium composition, around $Zn_{32}Cu_{54}Al_{14}$(B), at the annealing temperature $T_{annealed}$ (Fig. 1B-c). Based on the initial Zn/CuAl volume ratio, the maximum dissolution depth is estimated at 4.5 mm. This boundary is visible and produces a curved interface with the remaining CuAl, corresponding to the liquid meniscus at equilibrium ($H_{liqui}$d). The solidification front then progresses, driven by inter-diffusion, leading to a gradual depletion of Zn and enrichment of CuAl in the liquid. Solidification proceeds in a manner equivalent to a Bridgman process at an estimated rate of 1 mm/h, slower than the 20 mm/h used for standard growth of CuZnAl single crystals [25]. To determine the phases and compositions during solidification, the annealing was interrupted at different times, Figure1a, Figure 1d: it confirms the progressive solidification from liquid + β to β without intermediate phases.
As shown in Figure 1d, a sample air cooled to allow the stables phases to grow before the end of solidification exhibits the following phases: from top to bottom, a gray multiphase zone corresponding to the liquid plus β phase, and a single crystal developing from the initial meniscus. In the remaining 7 mm of solid $Cu_{79}Al$, interdiffusion ($D_{Zn}$ = $7.10^{-7}$ $cm^2/s$) takes approximately 100 hours to establish a continuous composition gradient. This gradient promotes diffusion induce re-crystallization (DIR) mechanism, in which the existing single crystal acts as a seed [26]. This synergy between directional solidification and solid-state diffusion is key to producing large single crystals with a composition gradient.

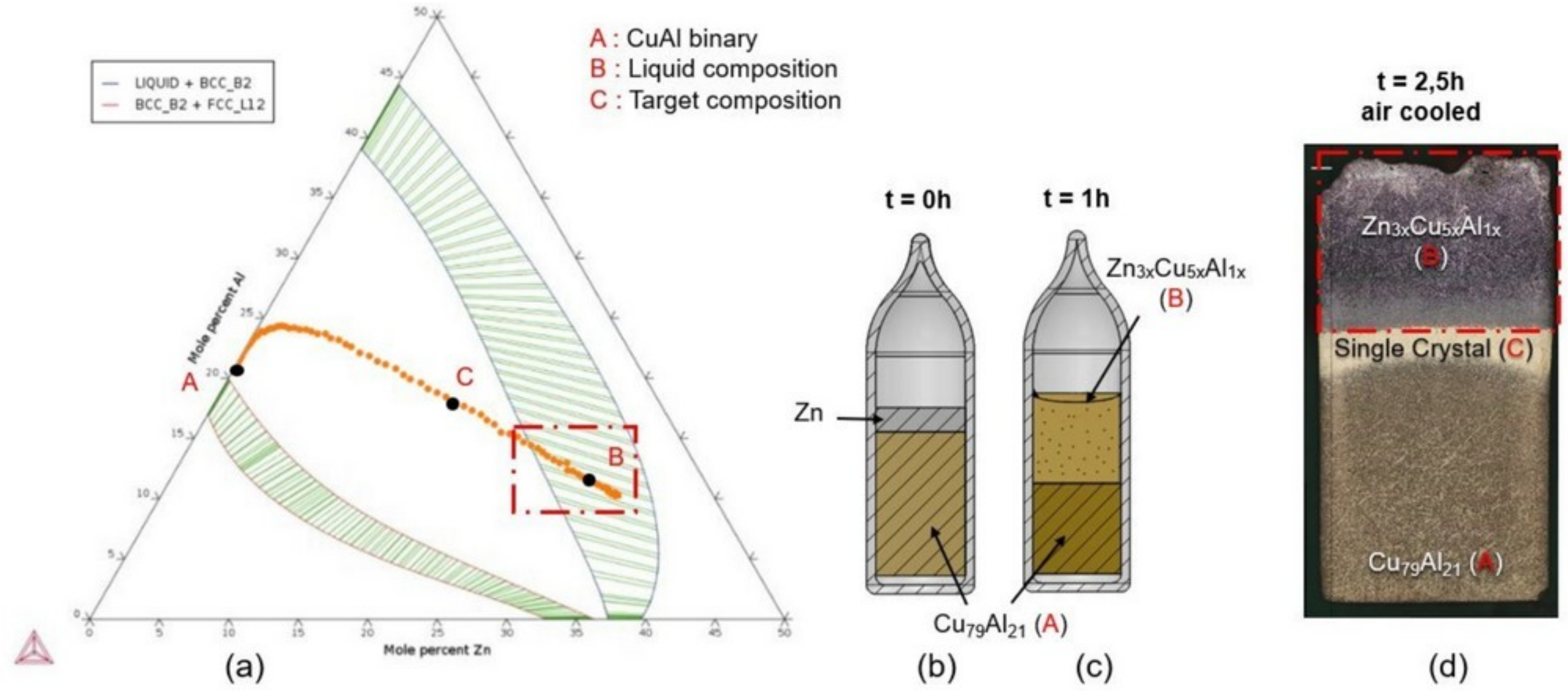

*Figure 1: (a) Isothermal section of the Cu-Zn-Al ternary phase diagram at 1173 K, showing the target composition (C) and the liquid composition (B). The orange curve (A–C–B) represents EPMA measurements along a sample quenched after 4 h. (b, c) Schematics of the process evolution during isothermal annealing. (d) Micrograph of a sample air-cooled after 2.5 h of treatment, exhibiting undissolved CuAl at the bottom, a single crystal in the middle, and the solidified liquid zone at the top (highlighted by the red box, corresponding to composition B in (a), which transforms via L+β→γ+β upon cooling)*

*Table 1:*

| Sample | Method | Average composition | ∇ Al (at.%/mm) | ∇ Zn (at.%/mm) | Length (mm) |
|---|---|---|---|---|---|
| Grad.1 | Liquid diffusion | $CuAl_{17.2}Zn_{13.8}$ | 0.13 | 0.11 | 11 |
| Grad.2 | Liquid diffusion | $CuAl_{16.2}Zn_{14.2}$ | 0.59 | 0.89 | 6 |
| Grad.3 | Liquid diffusion | $CuAl_{14.9}Zn_{16.9}$ | 0.44 | 0.36 | 9.5 |
| Grad.4 | Diffusion bonding | $CuAl_{14.6}Zn_{17.4}$ | 0.11 | 0.20 | 10 |

A single crystalline sample (Grad.1) with a gradient in chemical composition processed with this method is shown in Figure 2, the gradient reached around 1.3 at % $cm^{-1}$ for Zn, 1.1 at%/$cm^{-1}$ for Al, 0,2 at%/$cm^{-1}$ for Cu. EBSD analyses were carried out on a ZEISS Gemini 500 scanning electron microscope (SEM) along the compositional gradient, using a 10 µm probe. These reveal that the crystal is divided by a very low-angle sub-grain boundary (less than 1°) into two zones of approximately 4 mm, each exhibiting misorientations at the limit of the instrument’s resolution of 0.5°. Furthermore, Laue micro-diffraction patterns were obtained at ESRF on BM32 beamline at room temperature along the composition gradient. It reveals that between subgrains the misorientation reaches 0.15° and displays very well defined and sharp peaks. Indeed we analyzed the Laue patterns collected on maps distributed over the sample using LaueNN software [25] (50×50 µm mapping size with a step of 1 µm and a beamsize of 0.4 µm): the magnitude of all the deviatoric components of the elastic strain tensor are found to be less than 0.03%.

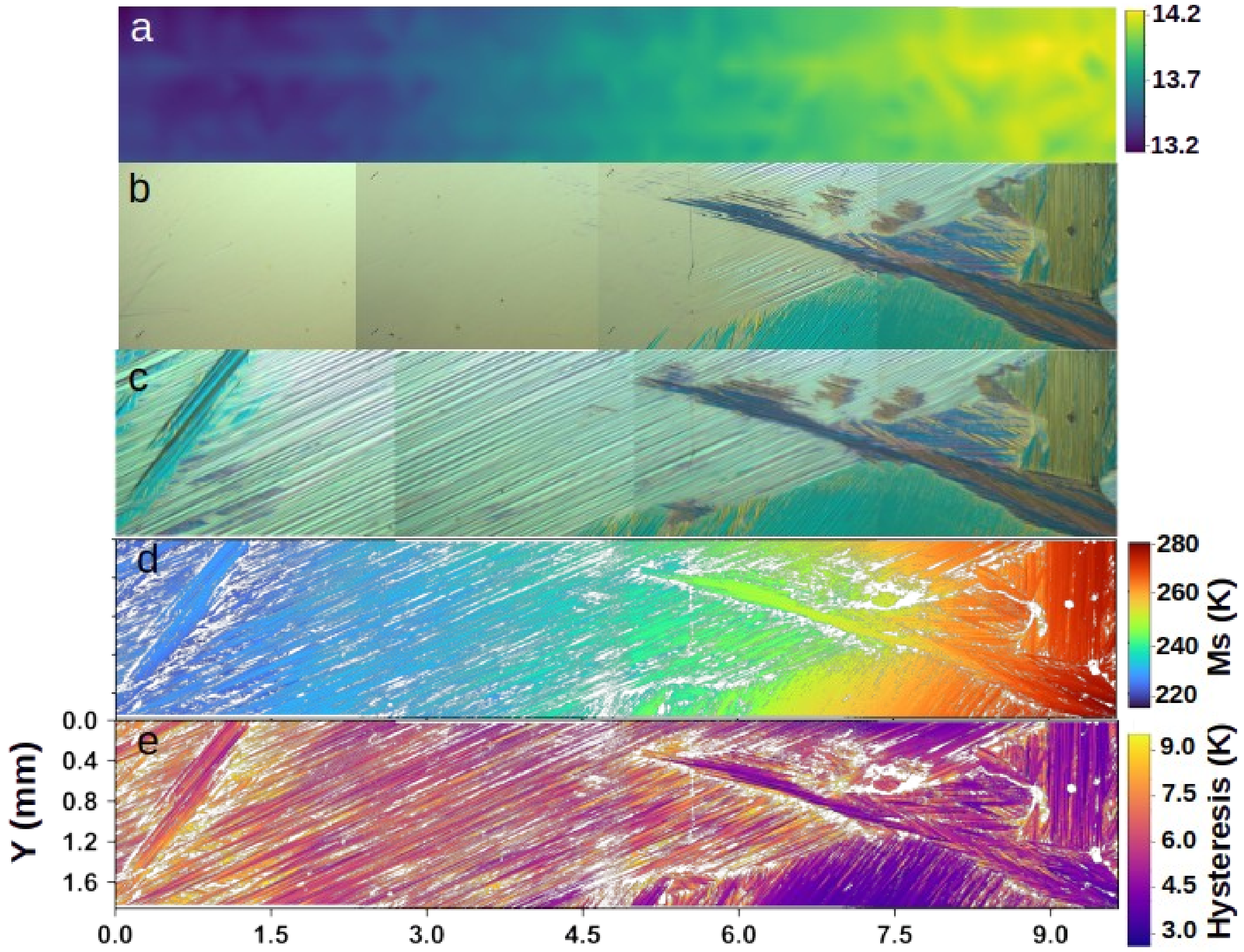

*Figure 2: (a) EPMA map of Zn content, (b) polarized light microscopy during cooling from RT: at 243K: austenite remains on the left and martensite microstructure front is on the right, (c) polarized light microscopy at 220K with full martensite phase. Corresponding cartography measured by PLM of local Ms (d), hysteresis (e)*

As a final step for all samples, to complete austenite ordering and removal of vacancies, annealing at 373K during 24h was systematically applied on all samples [27].

**Polarised Light Microscopy processing**

The evolution of surface relief associated with martensite forward and reverse transformations during thermal cycling was monitored using reflected light microscopy with polarized light. Observations were performed using a Zeiss AXIO optical microscope equipped with a ×5 objective, providing a field of view of approximately 2.7 × 2.25 mm, acquired with a CMOS camera (2464 × 2056 pixels, 8 bits per channel). Pixels were binned 4×4 (by summation), resulting in a resolution of 616×514 (16bits per channel). This operation reduce the number of pixels to be processed and therefore the computational time, while increasing the signal-to noise ratio (SNR). Synchronous temperature acquisition were performed automatically. An example of the fully developed martensitic microstructure in the Grad.1 concentration gradient at 233K is shown in Figure 2.

The samples were mounted using a thin layer of Apiezon grease on a 22W three-stage thermoelectric module capable of controlling the temperature between 373K and 198K. The cooling rate was set to 10 $K.min^{-1}$ using PID regulation (Meersteter engineering). To ensure complete transformation, the samples were cooled to 10 K below Mf and heated to 10 K above Af. Images were acquired at a rate of 2 frames per second, synchronized with temperature measurement, corresponding to a temperature increment of 0.083 K per frame.

The phase transformation from a uniform single-crystal austenite surface to a microstructure of martensite variants induce changes in surface topology that result in optical color changes using polarized light. The pixel intensity evolution during cooling/heating across the RGB channels can be monitored to track the phase transformation.

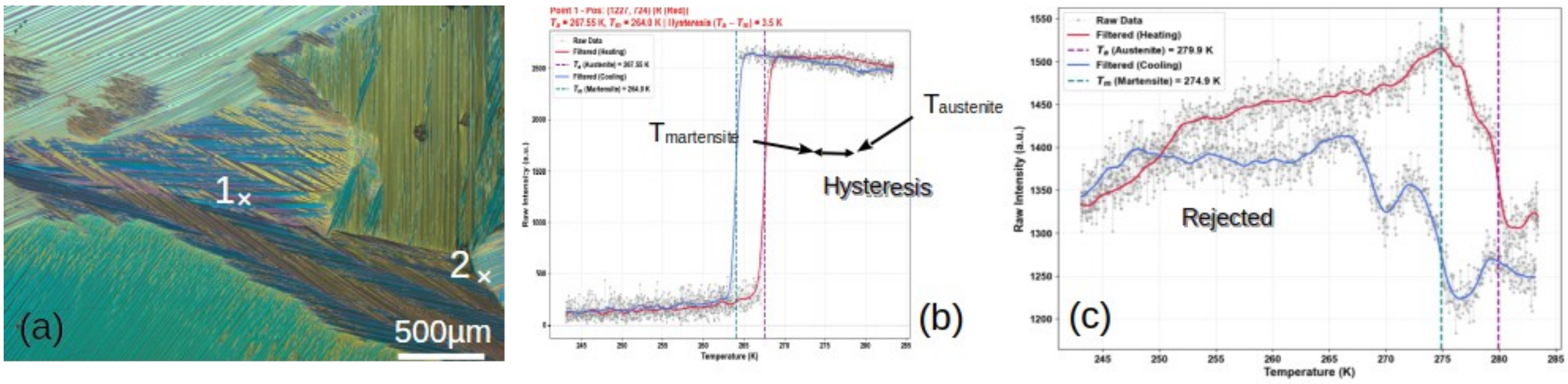

*Figure 3: (a) Micrography of sample 1,(b) and (c) cooling/heating curves of pixel 1 and 2 (reported on a)*

For each pixel in the recorded movie of heating/cooling sequence, its transitions temperatures were determined:

- The RGB channel exhibiting the highest variation in amplitude is selected, normalized and smoothed using a Gaussian filter ($\sigma$=10), expressed as I(T) (Figure 3b, c).

- From the derivative (dI/dT), the highest value during cooling and heating cycles were defined respectively as Tm and Ta, representing the characteristic temperature of austenite and martensite (Figure 3b).

- Because of some ill-defined transitions, (Figure 3c) which are mainly caused by surface defects or martensite variants exhibiting low variation relative to the austenite, a filtering step is performed based on Pearson’s coefficient ($R^2$>0.9) compared to a step transformation and a SNR > 5. These measurements are rejected and appear in white on hysteresis and Tm maps. This filtering resulted in an acceptation rate between 60-70% depending on the sample.

The local hysteresis width is calculated as the temperature difference $\Delta T_{hysteresis}=T_a-T_m$ (Figure 3b). Characteristic temperatures $T_m$ and $\Delta T_{hysteresis}$ are then plotted at their corresponding spatial coordinates yielding transformations heat-maps (Figure 2d, e).

When applied to composition gradient samples we correlate the thermal data with the local chemistry. The discrete composition field C(x, y) was linearly interpolated between neighboring grid points to generate a continuous function. Micro-indents served as fiducial markers to spatially align the composition field with the local hysteresis map, ΔThyst(x, y), and the martensitic transformation temperature map, Tm(x, y).

## Results and discussion

### Validation of PLM

In a first series of experiments, we compared PLM with DSC measurements (TA Instr. Q200) on homogeneous single crystal samples (±0.15 at %). The samples were cut so that each weighed between 10 and 20 mg (a disk 3 mm in diameter and 0.4 mm thick). For both methods, sample was subjected to 10K/min heating/cooling rate. A thin layer of thermal grease was applied [28] between the cell and the sample as bad thermal contact can induce lag in temperatures measurements [28]. To assess the ability of the PLM procedure to detect a phase change, a droplet of DI water was cooled at 250 K. The melting was recorded at a heating rate of 10 K.min-1. S. After applying the same image processing to these images, the melting point was determined to be 273.1 K.

In order to compare the two methods, the transformed martensite fraction was calculated from the DSC signal by integrating the heat flow over temperature, whereas for the PLM method, it was defined as the cumulative normalized number of pixels transforming at the surface (Figure 4a). The superposition of both curve is very close, however the end of the martensitic transformation finish earlier on the PLM signal (by ≈ 0.5K). This minor discrepancy arises because both methods are intrinsically different and cannot produce identical measurements: DSC is a bulk technique that measures heat flow through a calibrated thermal resistance, whereas PLM is restricted to the free surface, where the absence of three-dimensional elastic constraints allows the transformation to complete slightly earlier than in the bulk.

Theses curves can later be use to determine the global characteristic temperature (Ms, Mf, As, Af) of the sample by the tangent method. The hysteresis is measured by the area between the heating and cooling curves.

To ensure that the phase transformation behavior is uniform throughout the entire specimen, both sides of several samples were measured using the PLM method.

Variants of martensite run across the surface, displaying a similar pattern on both sides.

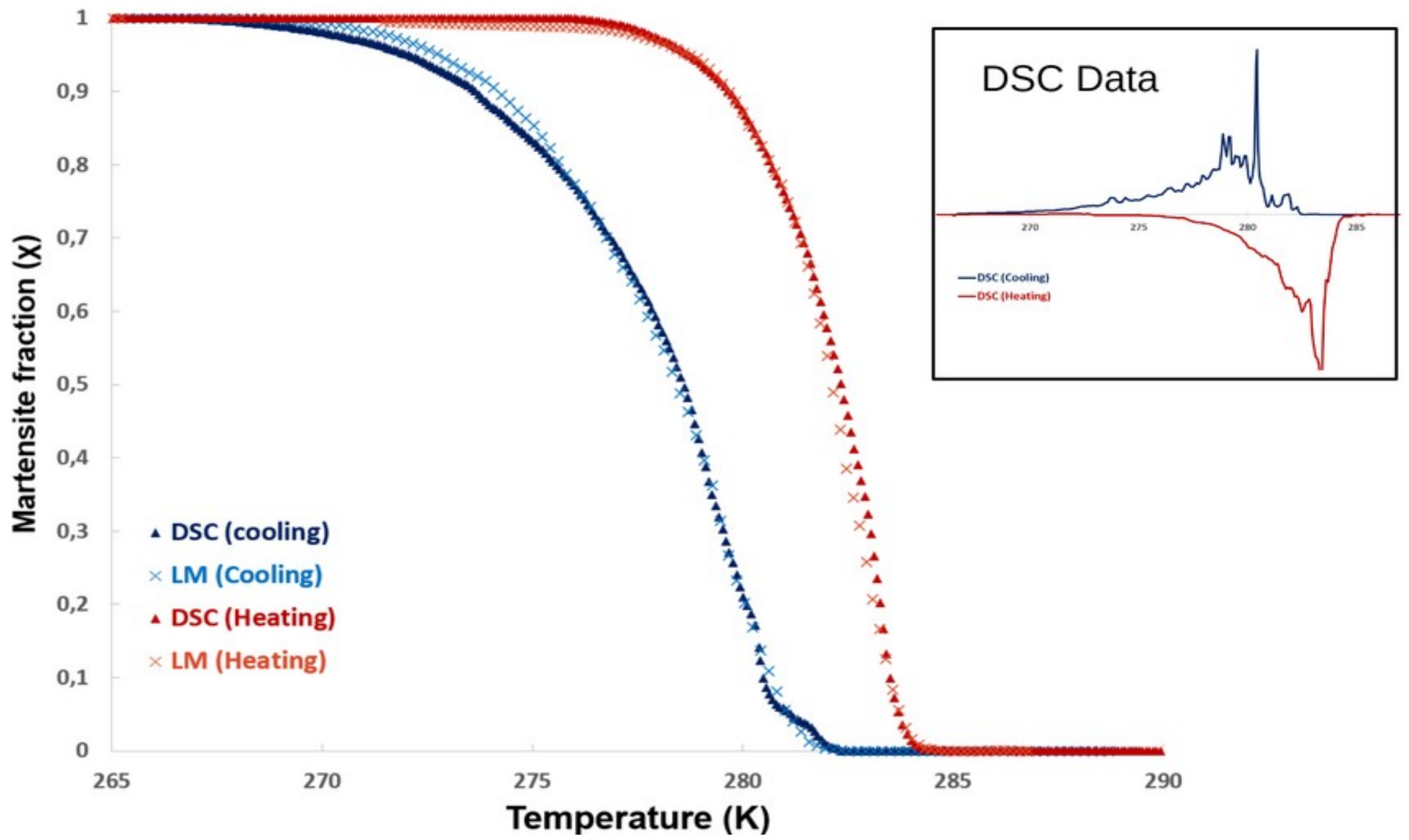

*Figure 4: (a) DSC heat flow and PLM surface flow normalized, (b) martensite fractions (surface,volume) obtained with the two methods*

Results of 4 samples of different compositions are summarized in 2 and show a maximum difference in Ms of 0.6K between the two methods.

Ultimately, the PLM method captures variations at the micrometric scale without inherent limit in sample size. This establishes it as a relevant technique for phase transition observations with spatially resolved microstructure evolution [19]

| | $Ms_{DSC}$ (K) | $Ms_{PLM}$ (K) | $Af_{DSC}$ (K) | $Af_{PLM}$ (K) | $Hysteresis_{DSC}$ (K) | $Hysteresis_{PLM}$ (K) |
|---|---|---|---|---|---|---|
| $CuZn_{16.8}Al_{14.8}$ | 286.0 | 286.6 | 292.7 | 292.3 | 7.1 | 6.9 |
| $CuZn_{20.0}Al_{12.9}$ | 226.3 | 226.5 | 228.8 | 228.7 | 2.7 | 2.2 |
| $CuZn_{11.1}Al_{19.0}$ | 281.1 | 281.3 | 283.7 | 283.9 | 4.1 | 4.1 |
| $CuZn_{22.0}Al_{11.0}$ | 238.6 | 238.9 | 241.9 | 242.0 | 4.2 | 4.3 |

*Table 2: Characteristic transformation temperatures (Ms, Af) and thermal hysteresis obtained via the tangent and integration methods from DSC and PLM signals for single crystals of different compositions*

Applying both methods across a wide composition range validates the applicability of the optical technique over the entire investigated domain.

**Martensite start**

The Ms temperature is plotted in ternary composition (at.%) for all composition gradients (Figure 5) and 16 homogeneous single crystals within a margin of error of +/-0.15at.%. Gradients (1), (2) and (3) were produced with the method described earlier, while (4) was processed by diffusion bonding. These composition gradients show a continuous and steep variation of the transformation temperature along the sample, with gradient (3) exhibiting a range of 80 K.
In contrast, a constant Ms line is measured in gradient (4) along its composition gradient. The resulting variation is restricted to ≈10 K for a 1.5 at.% variation in Zn, whereas gradient (1) shows a much steeper variation of 40 K for only a 1 at.% change.

While only a limited number of compositions in the literature have been measured with an accuracy comparable to our study (≈0.1 at.%), achieving such precision does not eliminate the scatter in transformation temperatures. In our experimental data, the spread of Ms, for equivalent compositions, still reaches an average of ±20 K (with one exception at +60 K). This scatter, which is widely observed in the literature [29], is an intrinsic characteristic of the CuZnAl system, where even sub-atomic percent variations within the measurement error drastically affect Ms. Furthermore, it is worth noting that the Ms values for composition gradients 3 and 4, which are superimposed, are all below 20K.

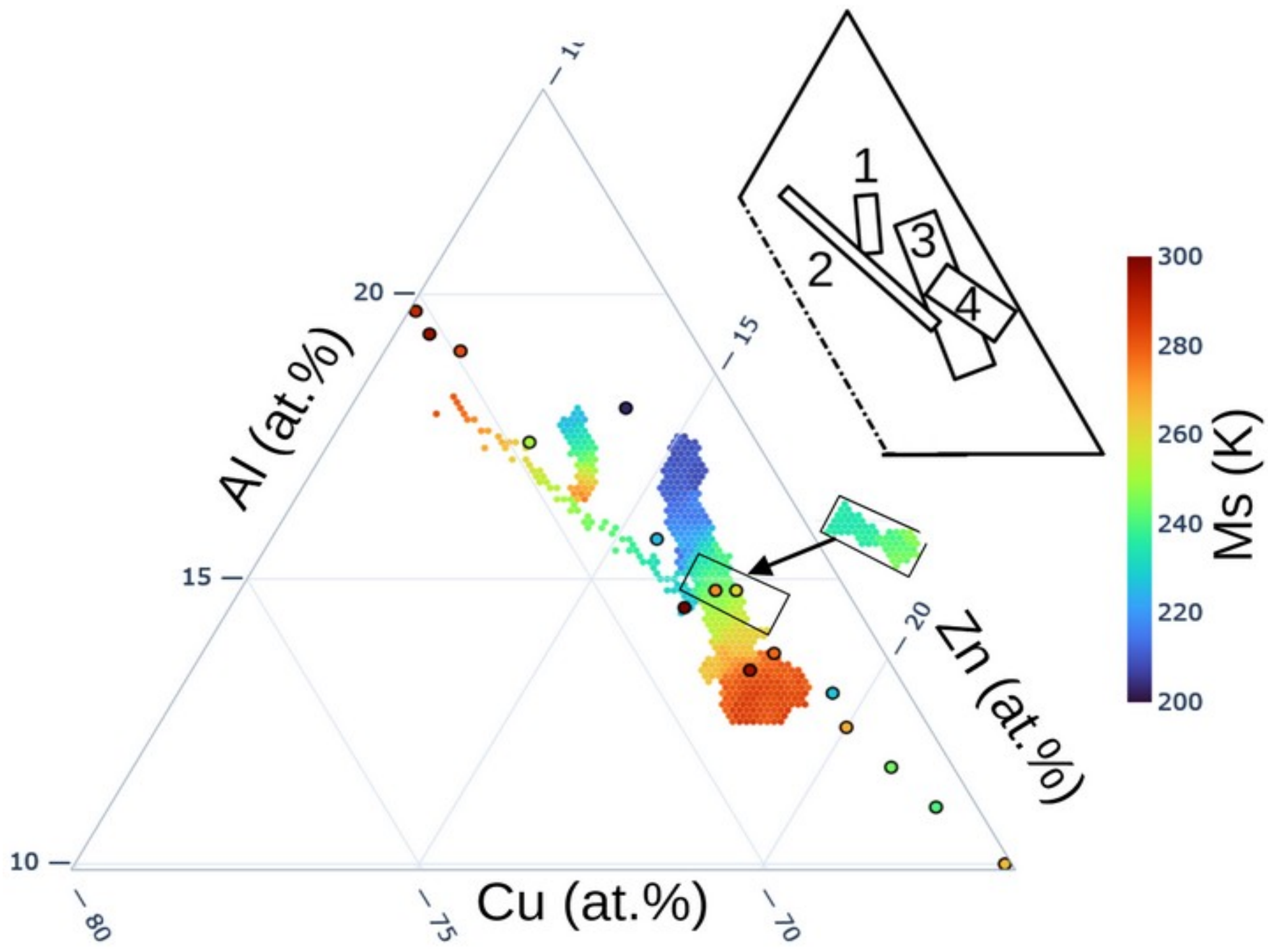

*Figure 5: CuZnAl diagram with, martensite start temperature on four composition gradients samples, single crystals with homogeneous compositions are also represented.*

Following the litterature, an empirical equation was derived from these data to estimate Ms as a function of Al and Zn concentrations in at/% ($C_{Al}$ and $C_{Zn}$ respectively):

$$Ms\ (K) = 1677 - 52.6 \times C_{Al} - 38.5 \times C_{Zn} \quad (1)$$

$$Ms\ (K) = 3280 - 110 \times C_{Al} - 80 \times C_{Zn} \quad \text{from [21]} \quad (2)$$

$$Ms\ (K) = 2309 - 74.5 \times C_{Al} - 54.7 \times C_{Zn} \quad \text{adapted from [30]} \quad (3)$$

$$Ms\ (K) = 2514.6 - 79.4 \times C_{Al} - 62.5 \times C_{Zn} - 12.5 \times C_{Ni} \quad \text{adapted from [31]} \quad (4)$$

While the established formulas (2,3) from [21] and [30] are widely used to estimate Ms, they predict a highly abrupt behavior outside of the standard composition range where small compositional changes result in extreme Ms variations. Such drastic variations were not observed in our samples. For instance, using equ. (2) , the copper-rich side of our gradient was predicted to have a Ms of 400 K, while the copper-poor side was predicted at 150 K. Equation (3) exhibits a similar, though less abrupt, trend.

Despite these discrepancies in absolute sensitivity, the models generally agree on the proportional impact of the alloying elements. In our newly formulated equation, the ratio of the relative effect of Al and Zn on Ms is 1.362, compared to 1.375 for equ. (2). It should be noted that equ. (3) was initially constructed using a combination of weight percent (wt.%) and electron-to-atom ratio (e/a); therefore, its conversion to atomic percent (at.%) is an approximation. Taking this into consideration, the Al/Zn effect ratio derived from equ. (3) is also estimated at 1.366. Finally [31] recently proposed equation (4) including the effect of Ni, which gives a good overall estimation (close to equ. (3)) but the ratio of Al/Zn is lower (1.27). Ultimately, all formulas are in very good agreement regarding the stronger relative influence of Al compared to Zn.

**Hysteresis measurements**

During initial repeatability tests, the martensitic microstructure varied from cycle to cycle before stabilizing into a reproducible pattern.

The local hysteresis values are plotted on the ternary diagram (Figure 7), 0.1 at.% points for local composition map, 0.3 at.% for homogeneous samples. In the assessed area, hysteresis varies between 2 and 8 K, with a bell-shaped curve distribution along each sample with a standard deviation between 1.5 K and 1.9 K. The average hysteresis is 4.1 K for Gradients (3) and (4), 4.5 K for Gradient (2), and 6.1 K for Gradient (1).

Gradient (3) appears to be divided into three areas of constant hysteresis extending over 1 at.% of Zn. With increasing zinc content, these regions show average hysteresis values of 3.5 K, 4.5 K, and 3.5 K, respectively, a mere variation of 1K. Gradients (2) and (4) intersect gradient (3) in its central region and demonstrate a good agreement, the decrease in hysteresis with decreasing Zn content is also observed for gradient (2).
Gradient (1) and (2) both extend toward an Al-rich region where both show a localized increase in hysteresis, reaching a maximum of 8 K within a narrow, overlapping composition range. At this stage, to establish a clear correlation between composition and hysteresis for these composition regions, it would be necessary, on the one hand, for the measurements to be reproduced on a larger portion of the ternary system, and, on the other hand, to verify that the increase in hysteresis is not related to a structural change in the monoclinic structure (particularly toward aluminum-rich compositions with 2H structures [35]).

Homogeneous single crystals exhibit a range of variation similar to that of gradient samples, ranging from 2.7 K to 5 K. Here again, the discrete sampling in composition prevents the observation of a clear correlation between hysteresis and composition. Nevertheless, it is worth noting that the low hysteresis values (2.7–5.0 K) are consistent with those reported by Song et al. (2.0–5.1 K) over a narrow composition range in the analogous $Zn_{45}Au_xCu_{55-x}$ system with M18R transformation, adjusted to be close to geometric compatibility [9].

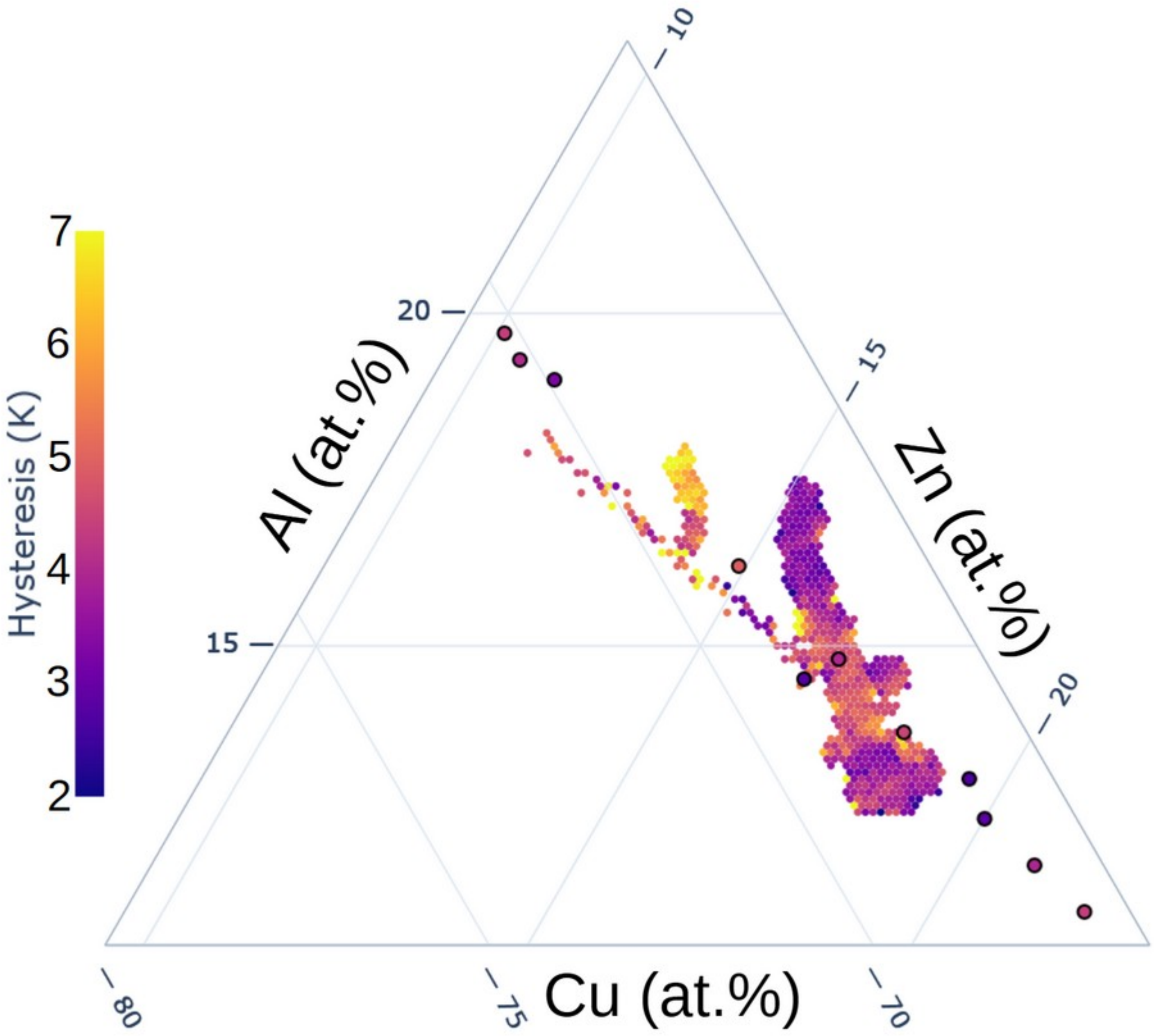

*Figure 6: (a) CuZnAl diagram with hysteresis width of composition gradients and homogeneous samples.*

**Microstructure**

The phase transformation can be described by shear of the cubic austenite lattice that lead to 24 possible monoclinic martensite variants. The CuZnAl variants form 6 equivalent self accommodating groups of 4 variants (A,B,C,D) with respect to a (011) plane [32]. They form typical wedge-shaped, interlocking plate microstructures [32, 36]. Variants have twin orientation relationship (type I,II,compounds, see table 6 in [11] for example). The microstructure is self-adaptative and multiscale: starting from the long-period stacking order (LPSO) of the martensite unit cell itself, TEM reveal the stacking-fault substructure inside each plate, then the plates themselves, and up to the mesoscale plate groups and wedges observed optically – the whole arrangement minimizing the total energy. Thus, TEM and optical observations show ranges of plate width between ~50 nm up to 100 µms [36, 32]. There is a uniform and discrete distribution of colors within well-defined regions, bounded solely by straight interfaces in our PLM setup: it is assumed that the few discrete color distributions within the microstructure are associated with distinct wedge groups. During the transformation, wedge-shaped structures form at various scales with the wedge serving as a growth path that promotes martensitic transformation [11]. What is observed on the surface does not necessarily correspond to elementary variants, and each plate visible under PLM may exhibit fine internal twinning that lies below the optical resolution [15]. The colours in figure 7 therefore corresponds to plate groups (internally twinned variants) rather than to strictly individual variants.

23 distinct combinations of variants can be classified into four families of twin relationships (reflection twins, 180° rotation twins, 120° rotation twins and 90° rotation twins). [36] demonstrated, using transmission electron microscopy (TEM) and high-resolution electron microscopy (HREM), that the three types of boundaries within the same group are clearly distinguishable from one another. The A/C boundary is straight and perfectly coherent; the A/B boundary is irrational ($K_2$ plane), coherent, and slightly curved; the A/D boundary is stepped, with greater misalignment between the crystallographic planes of the two variants and the boundary where these last two meet. Similarly, boundaries between groups do not exhibit the same coherence and junctions parallel to A/B or A/C type planes can appear relatively straight and coherent while others can behave like grain boundaries; their energy is high because both self-accommodation (γ_self) and misorientation (γ_mis) contributions are unfavorable.

The energy of martensite-martensite boundary is characterized by the twin boundary energy γ_tw, defined as the surface energy of the twin plane [37, 36, 12], to which two excess contributions are added: γ_self, the boundary energy contribution from imperfect self-accommodating of the variant pair's transformation shear – i.e. residual elastic energy stored at the interface – and γ_mis the contribution from the misorientation of the matching planes. γ_self + γ_mis + γ_tw ranks the boundary type : A/C < A/B < A/D << intergroups. More generally, the local hysteresis variations reflect the total energy minimisation of the system, which involves a trade-off between the martensite-martensite interfacial energy and the elastic accommodation energy of the lattice misfit [38]. Although some variant pairs have a low γ_tw, they may require a high elastic energy to accommodate the transformation strain mismatch with the austenite matrix. Conversely, pairs with higher γ_tw can be favored if they allow a better elastic accommodation (lower elastic energy) leading to a lower total energy [37]. It should be noted that this near-compatibility is not tied to a fine-tuned composition : Balandraud et al. demonstrated that wide domains of lattice-parameter space yield almost compatible, low-stress microstructures, consistent witht the same wedge microstructure and similar local hysteresis behaviour observed across our whole composition gradients [Ballandraud, Zanzotto].

This robustness is rather surprising when contrasted with strict supercompatibility ($\lambda_2$ = 1), which occurs only in narrow composition wells [Cui].

The boundary hierarchy is directly visible in our PLM hysteresis map. Several types of combinations of wedge-shaped and lamellar variants in two different samples exhibiting a compositional gradient are shown in Fig. 7 and exhibit variations ranging from 3 to 8 K in correlation with the microstructure. The local hysteresis shown in Fig.7 is relatively low (between 3 and 5 K) and is distributed uniformly, such that the intervariant boundaries are barely distinguishable. We therefore attribute these boundaries to coherent, mobile, low-energy interfaces that store little elastic energy, such as the A/B and A/C boundaries described above. The local hysteresis shown in Fig. 7 can reach 8 to 9 K and could therefore be associated with incoherent A/D-type or inter-group boundaries. This is the direct experimental signature of quasi compatibility in 18R Cu-Zn-Al as predicted by [11]. The wedge/plate group microstructure is not kinematically compatible, so residual elastic stresses are stored at and around these imperfectly compatible interfaces and pinned, high energy boundaries dissipate energy friction during thermal cycling [12]. Figure 7c illustrates a third case in which a hysteresis gradient is observed within a single variant, which lessens as one moves away from the boundary of that variant. Two mechanisms come into play: (i) the residual stress field extends beyond the boundary and decreases toward the interior of the variant, altering the local transformation temperatures according to the Clausius-Clapeyron relationship $d\sigma/dT$; (ii) defects such as the internal stacking fault substructure and dislocations that could be introduced by training provide distributed pinning sites. Consistently, Lovey and Torra [39] measured a hysteresis of only 0.06K for a single isolated interface versus ~ 10K in bulk, showing that the excess dissipation of the bulk transformation comes from interactions between martensite domains and their interfaces. No hysteresis level is attributed here to a particular variant population since the variants are not crystallographically individually identified

Altogether, the local hysteresis therefore maps the local degree of compatibility : coherent, self-accommodated boundaries dissipate little, while incoherent, misoriented, non self-accommodated boundaries – and the stress fields they generate inside the variants – concentrate the dissipation. This is why Cu-Zn-Al 18R, which is quasi-compatible, exhibits a mean hysteresis of ~ 5K, whereas strictly supercompatible alloys ($\lambda_2$ = 1, cofactor conditions; [9]) such as monoclinic Au-Cu-Zn show a mean hysteresis of ~ 2K, with $\lambda_2$ = 1 where the energy barriers at the interface vanish. One prediction arising from this interpretation is that hysteresis mapping, performed using the same optical method on a single-crystal sample of supercompatible composition, should result in spatially uniform hysteresis. Note also that the composition gradient itself contributes negligibly at the scale of one map (~0.2 K over 0.65 mm at ~1 at%/10 mm), so the 2.9–9.2 K dispersion is essentially microstructural.

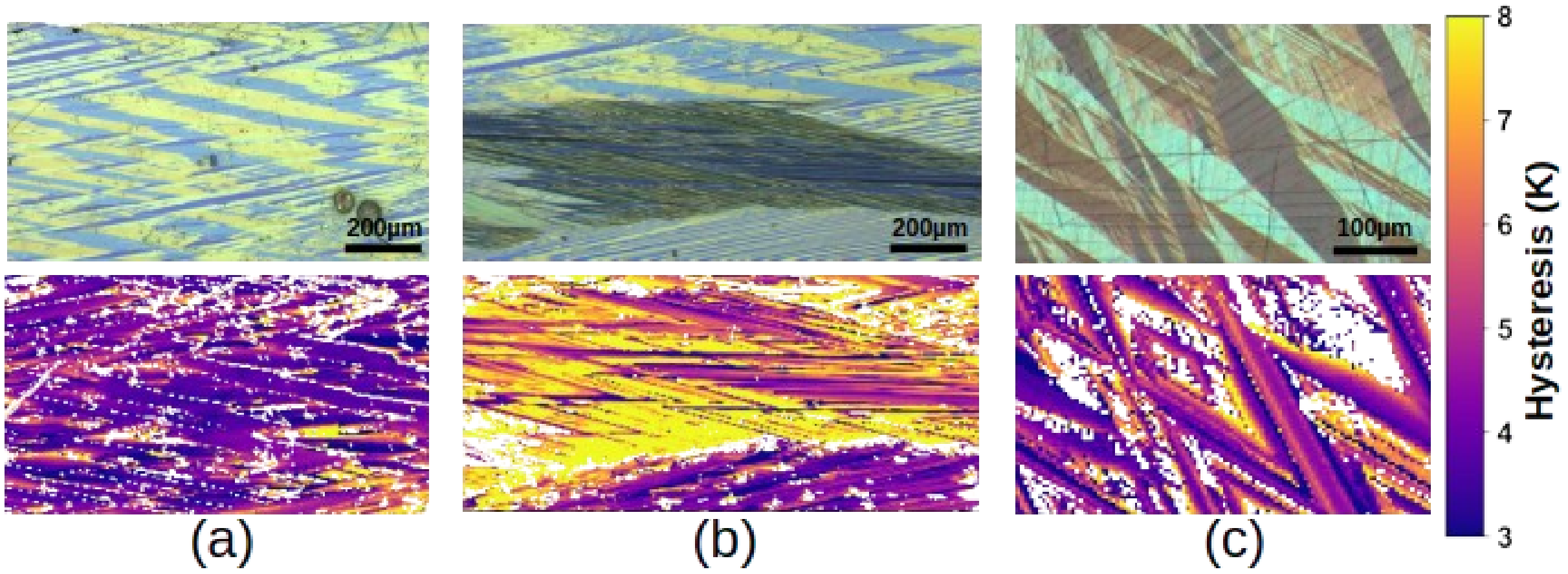

*Figure 7: Optical micrographs (top) and corresponding pixel-resolved thermal hysteresis maps (bottom) of the 18R martensite in the composition-gradient single crystal. (a) Coherent, self-accommodated intervariant boundaries (A/B, A/C type). (b) Incoherent, misoriented boundaries (A/D-type, inter-group). (c) Intra-variant hysteresis gradient decaying from the variant boundary.*

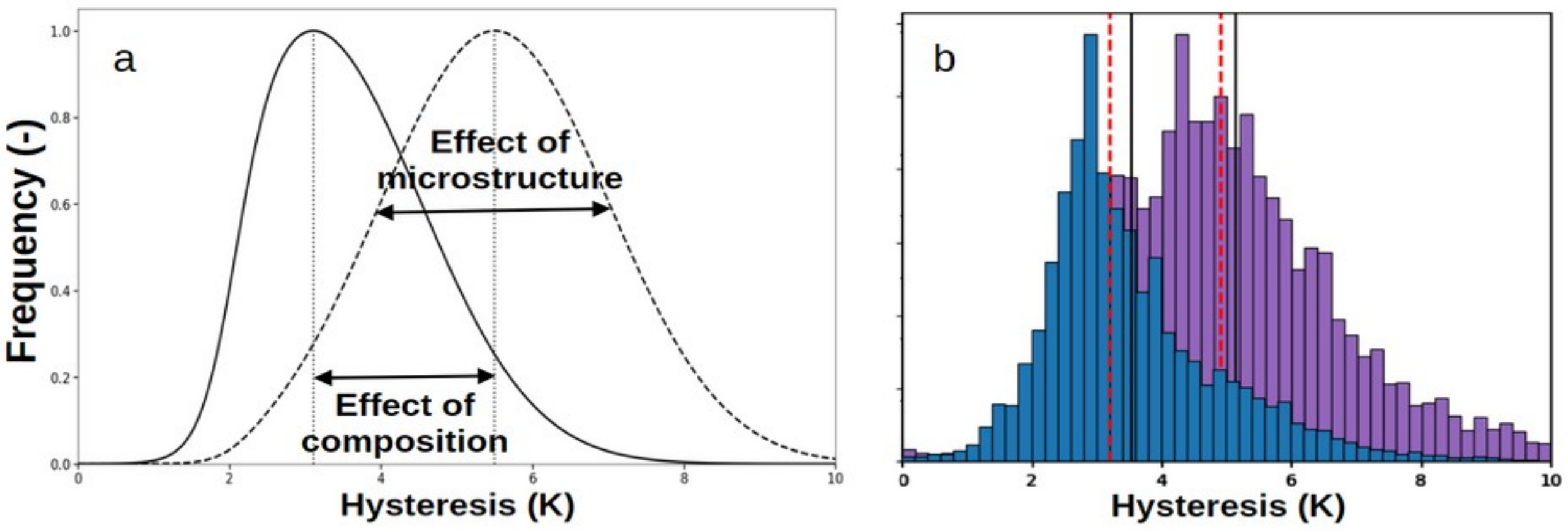

*Figure 8: (a) effect of both contributions, (b) hysteresis dispersion of two composition of Grad.3, (purple $CuAl_{14.2}Zn_{17.3}$, blue $CuAl_{16.4}Zn_{15.5}$), dashed line median, continuous line average.*

Composition and microstructure effects combine as follows: the composition sets the average hysteresis level of the sample, while the local variant population broadens the distribution of hysteresis values around this average (Figure 8). The lowest values, close to 3 K, are found in regions containing a single group, and the highest ones at the intersections of multiple variant colonies. Although this framework remains qualitative, it provides a useful basis for interpreting the experimental observations.

This method could be combined with full-field strain measurements during the transformation by localized spectrum analysis (LSA) of a laser-engraved periodic pattern [33, 34]. This gives access to the local displacement field, from which the deformation gradient of each variant colony can be derived, this yields the local transformation stretch matrix, which can be directly compared to the predictions of the crystallographic theory [33]. Alternatively, cryo-EBSD performed on the martensite provides the twin relationships between the variant populations and would bring high spatial resolution. Local measurements of crystal parameters (micro X-ray diffraction or micro-Laue) would give the elastic strains of the austenite and the stretch matrices characterizing the variant colonies. This ongoing work combines micro-Laue diffraction with local strain measurements. Combining PLM hysteresis maps with micro-Laue diffraction would allow direct measurement of the local elastic strain ans stretch matrices U_i of each colony, enabling quantitative évaluation of the compatibility criteria ($\lambda_2 \sim 1$) and the cofactors conditions $Fi-Fj = a \otimes n$ for each observed pair, thus linking the observed local hysteresis to the computed γ_tw and elastic accommodation energy of each variant pair [38, 9].

From an application perspective, these results are particularly relevant for elastocaloric devices, where transformation is typically driven under uni-axial loading conditions that favor the growth of a limited number of variants. Promoting the formation of variants with low hysteresis and favorable compatibility with the austenite could therefore improve functional stability and fatigue resistance.

## Conclusions

An experimental methodology to explore the effect of fine composition variations on the hysteresis width of the austenite-martensite transition in β-CuZnAl alloys is proposed in this work. The investigations support the following conclusions:

- A simple processing route is detailed for growing bulk single crystals in order to explore the composition range of beta brass CuZnAl with controlled composition gradients (down to 0.1 at%/mm). It is based on a combined diffusion driven Bridgman-like growth and diffusion induced re-crystallization. This method could be applied to other doped copper-based alloys. For a wide range of materials, diffusion bonding remains a very attractive alternative.
- All samples were produced, processed, and trained under identical conditions, in an effort to insure reproducible results.

- A local method, based on optical microscopy, has been developed to track the thermal hysteresis of phase transitions, enabling the evaluation of characteristic temperatures, and determine hysteresis with a higher spatial resolution than conventional techniques. Although there were differences in the cycle shape between PLM and DSC, hysteresis measured by area are equivalent.
- Within the range of compositions studied, it was found that the hysteresis is very low in CuZnAl (3-5K) and varies little, with composition.
- Local variations of hysteresis correlated to variant orientation and interactions with each other are observed and attributed to a variation of compatibility conditions with the parent phase or between martensites variants.

**Acknowledgments**

Authors kindly acknowledge Univ. Grenoble Alpes for providing an IRGA PhD grant for one of us (O.N). This research was funded by the French National Research Agency (ANR) as part of project ANR-25-CE50-2328-01 as well as Institute Carnot Energies for complementary funding (CASTLE project). We acknowledge the *European Synchrotron Radiation Facility* (ESRF) for provision of beam time on IF beamline (experiment A32-1-208 (2025). Fruitful discussions with A. Tolley, G. Castro (Bariloche) in the framework of ECOS Sud (2024) collaboration are also acknowledged.